\documentclass[a4paper]{jpconf}
\usepackage{graphicx}
\usepackage{multicol}
\usepackage{etoolbox}
\usepackage{relsize}
\begin{document}
\title{The COSINUS project - a NaI-based cryogenic calorimeter for direct dark matter detection}

\patchcmd{\thebibliography}
  {\list}
  {\begin{multicols}{2}\smaller\list}
  {}
  {}
\appto{\endthebibliography}{\end{multicols}}

\newcommand{\addrMPI}{$^1$}
\newcommand{\addrINFNMilano}{$^2$}
\newcommand{\addrUniMilano}{$^3$}
\newcommand{\addrOeAka}{$^4$}
\newcommand{\addrAtomI}{$^5$}
\newcommand{\addrGSSI}{$^6$}
\newcommand{\addrLNGS}{$^7$}

\author{G.~Angloher\addrMPI
        ,
        P.~Carniti\addrINFNMilano
        ,
        L.~Cassina\addrINFNMilano
        ,
        L.~Gironi\addrINFNMilano$^,$\addrUniMilano
		    ,
		    C.~Gotti\addrINFNMilano
		    ,
		    A.~G\"utlein\addrOeAka$^,$\addrAtomI
		    ,
		    D.~Hauff\addrMPI
		    ,
		    M.~Maino\addrINFNMilano
		    ,
		    S.S.~Nagorny\addrGSSI
		    ,
		    L.~Pagnanini\addrGSSI
		    ,
		    G.~Pessina\addrINFNMilano
		    , 
		    F.~Petricca\addrMPI
		    ,
        S.~Pirro\addrLNGS
		    ,
		    F.~Pr\"obst\addrMPI
		    ,
        F.~Reindl\addrMPI$^,$$^*$,
	 	    K.~Sch\"affner\addrGSSI$^,$\addrLNGS   
	 	    ,
	 	    J.~Schieck\addrOeAka$^,$\addrAtomI   
	 	    ,
	 	    W.~Seidel\addrMPI
}

\address{
    \addrMPI Max-Planck-Institut f\"ur Physik, M\"unchen - Germany, 
		\addrINFNMilano INFN - Sezione di Milano Bicocca, Milano - Italy, 
		\addrUniMilano Dipartimento di Fisica, Universit\`{a} di Milano-Bicocca, Milano - Italy, 
		\addrOeAka Institut f\"ur Hochenergiephysik der \"Osterreichischen Akademie der Wissenschaften, - Austria, 
		\addrAtomI Atominstitut, Vienna University of Technology,Wien - Austria, 
    \addrGSSI GSSI - Gran Sasso Science Institute, L'Aquila - Italy, 
    \addrLNGS INFN - Laboratori Nazionali del Gran Sasso, Assergi - Italy
  }

\ead{$^*$florian.reindl@mpp.mpg.de}

\begin{abstract}
  At present the results in the field of direct dark matter search are in tension: the positive claim of DAMA/LIBRA versus null results from other experiments. However, the comparison of the results of different experiments involves model dependencies, in particular because of the different target materials in use. The COSINUS R\&D project aims to operate NaI as a cryogenic calorimeter. Such a detector would not only allow for a direct comparison to DAMA/LIBRA, but would also provide a low(er) nuclear recoil threshold and particle discrimination. 
\end{abstract}

\begin{figure}[htb]
  \begin{minipage}[t]{0.49\textwidth}
    \centering
    \includegraphics[width=.80\textwidth]{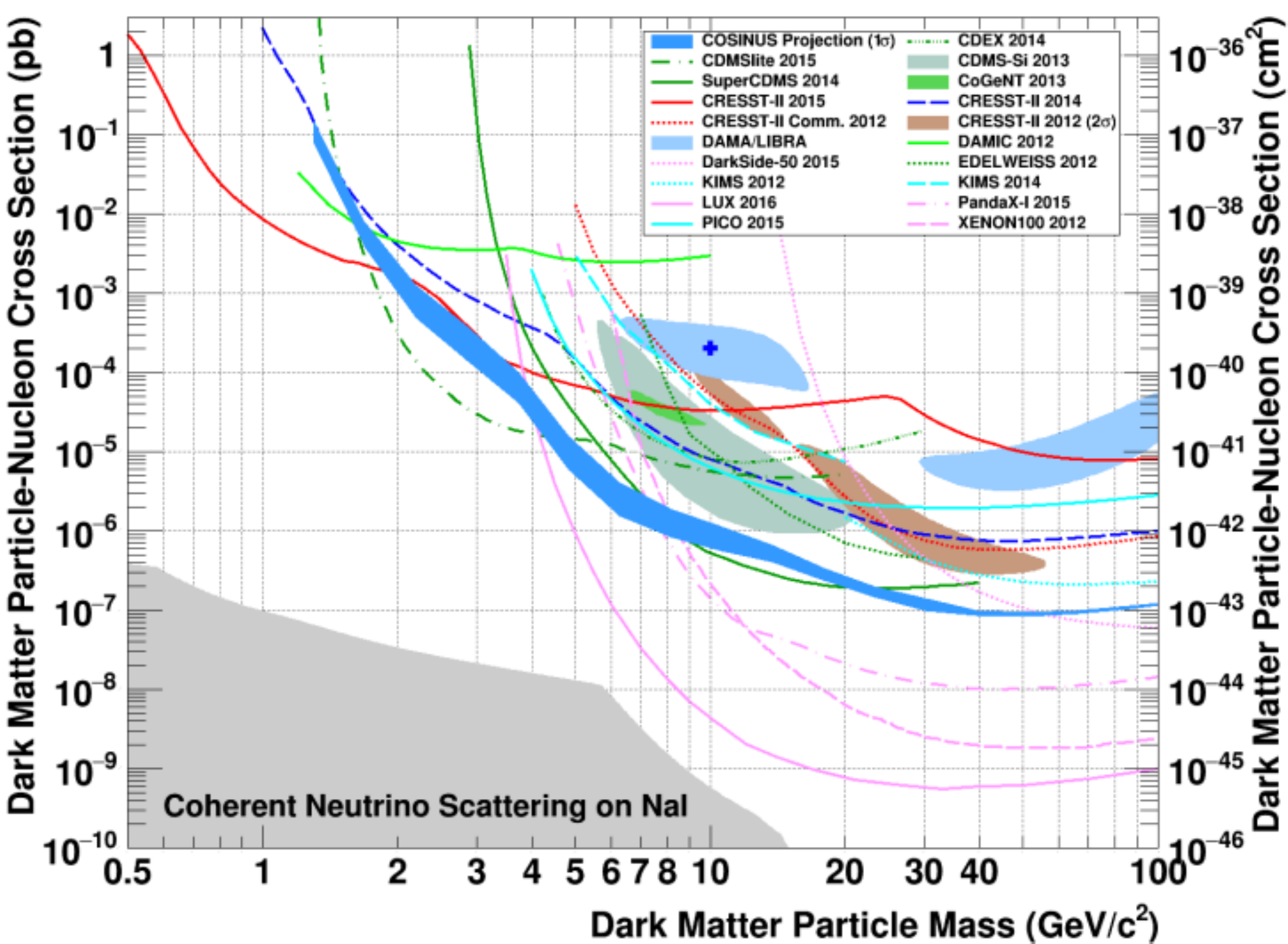}
    \caption{Anticipated sensitivity of COSINUS (blue band) for spin-independent elastic scattering. Also shown: the benchmark point (blue cross, see text) and results from other experiments. Plot from \cite{angloher_results_2016}, references given therein.}
  \label{fig:limit}
\end{minipage}
\hfill
\begin{minipage}[t]{0.49\textwidth}
    \centering
    \includegraphics[width=.80\textwidth]{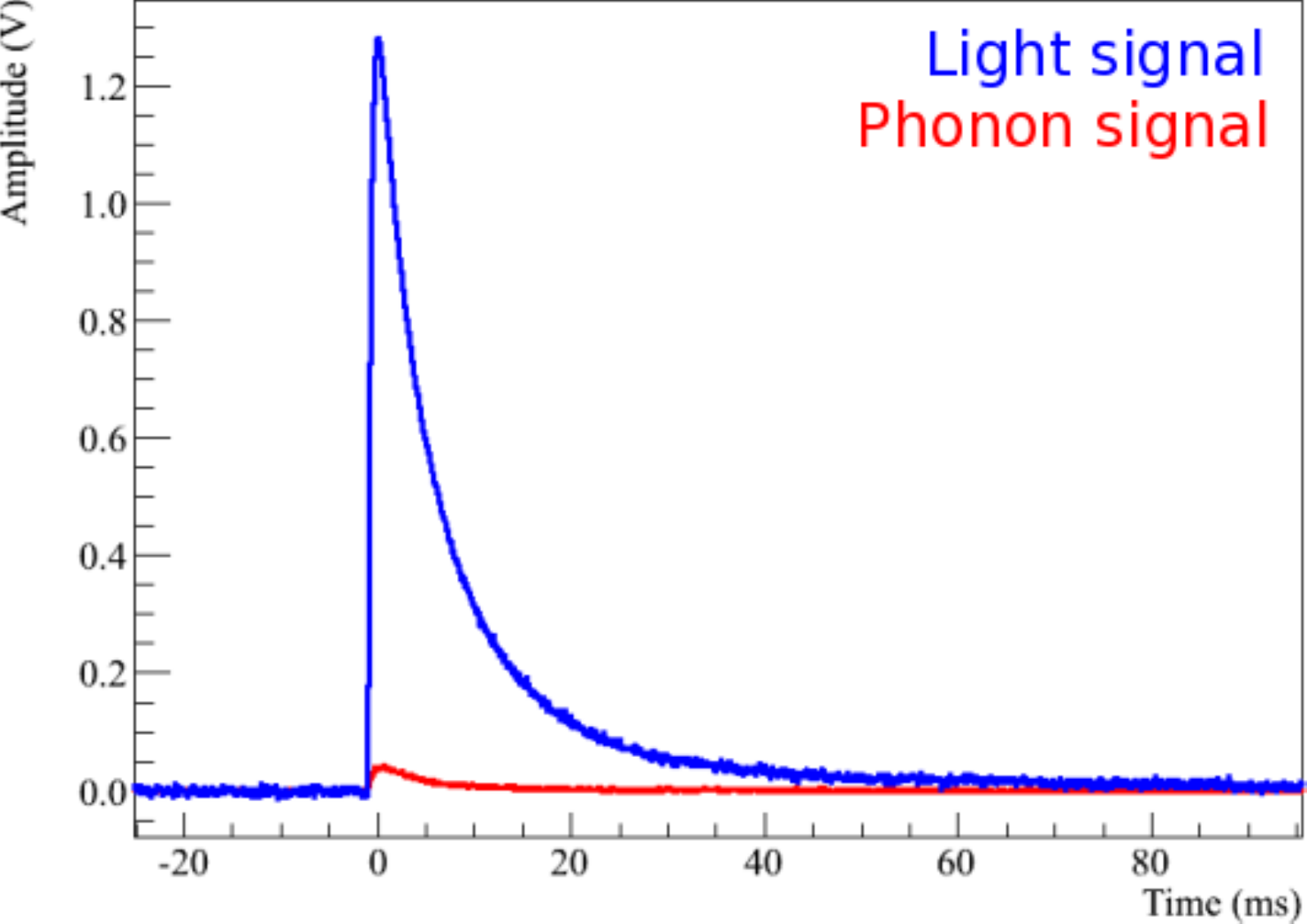}
  \caption{e$^-$/$\gamma$-event with an energy of 240 keV measured with the first NaI prototype. Simultaneously recorded are the light signal (blue) and the phonon signal (red).}
  \label{fig:Pulse}
\end{minipage}
\end{figure}

Today, the existence of dark matter (DM) is certain, its underlying nature, however, is still largely unknown. %
At present, the DAMA/LIBRA experiment\footnote{DAMA/LIBRA will be shortened to DAMA in the following.} claims to observe dark matter via its expected annual modulation \cite{bernabei_final_2013}, which disagrees with null results from many other direct DM searches (fig.~\ref{fig:limit}, \cite{savage_compatibility_2009}). Stating disagreement, however, only holds for certain assumption on the interaction mechanism of DM with Standard Model particles. The major model-dependence in the comparison of experiments comes from the use of different target materials: DAMA uses NaI(Tl), while the excluding null results are obtained with Ar, CaWO$_4$, CsI, Ge, Si and Xe targets. 

The R\&D project COSINUS\footnote{Cryogenic Observatory for SIgnals seen in Next-generation Underground Searches} aims to develop a NaI-based cryogenic detector \cite{angloher_cosinus_2016} offering particle discrimination and, as also using NaI, its results can be directly compared to DAMA. 

The heart of a COSINUS detector is an undoped NaI crystal, cooled to mK temperatures. Any particle interaction in the crystal will excite phonons. Recording this phonon signal with a so-called transition edge sensor (TES) provides a very precise measurement of the energy deposited in the crystal, quasi independent of the type of interacting particle. In addition, also scintillation light is produced by a particle interaction which we measure by pairing the NaI crystal with a cryogenic light detector, read-out with a second TES\footnote{The TESs are produced by the Max-Planck-Institute of Physics in Munich.}. Since the amount of produced scintillation light strongly depends on the interacting particle, measuring phonons and light provides particle identification. In particular, it allows to discriminate e$^-$/$\gamma$-events from nuclear recoils; the former being the main background, the latter being believed (in the vast majority of models) to be induced by DM.

\begin{figure}[htb]
  \begin{minipage}[t]{0.49\textwidth}
    \centering
    \includegraphics[width=.80\textwidth]{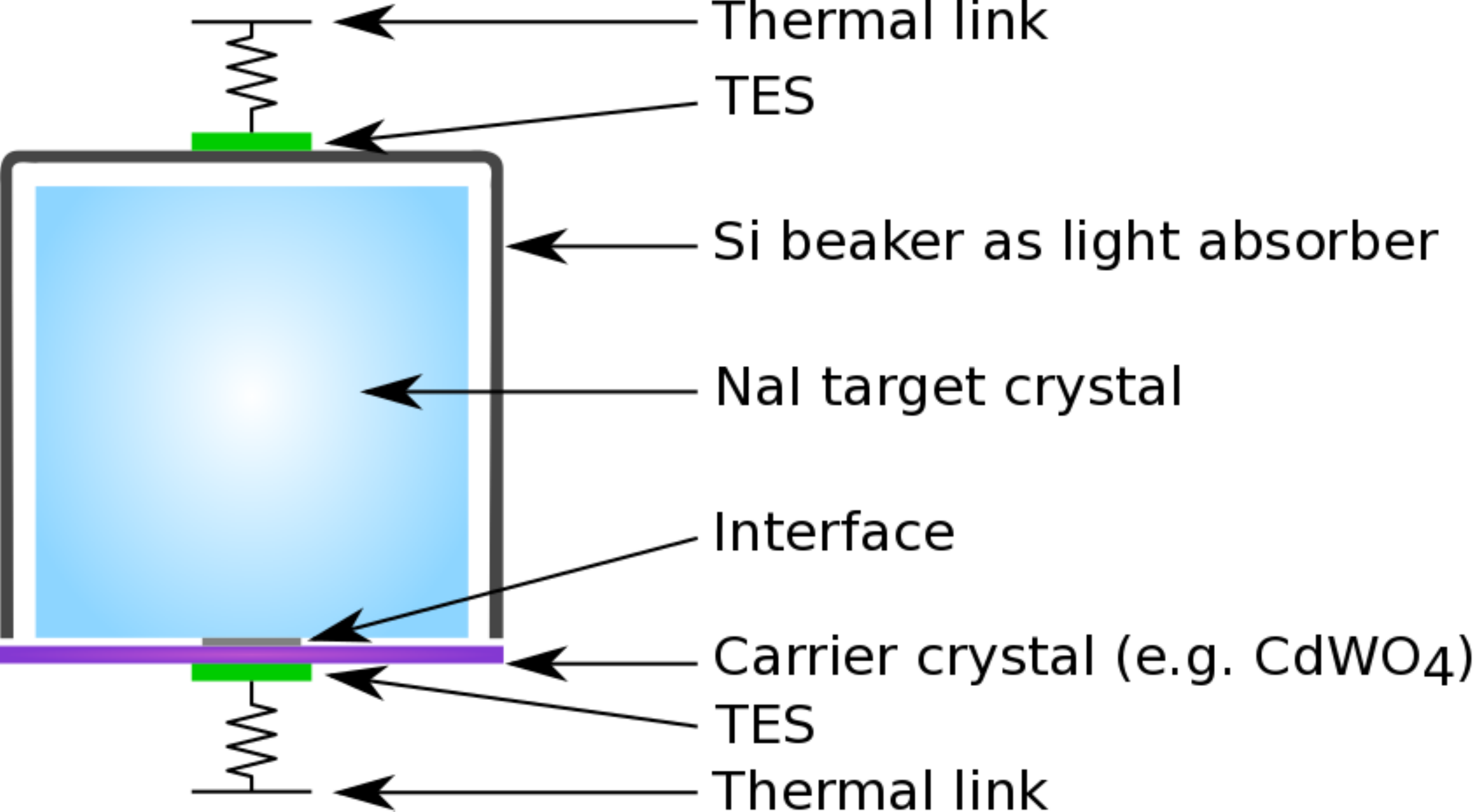}
    \caption{Schematic of a COSINUS detector.}
    \label{fig:schema}
  \end{minipage}
  \hfill
  \begin{minipage}[t]{0.49\textwidth}
    \centering
    \includegraphics[width=.80\textwidth]{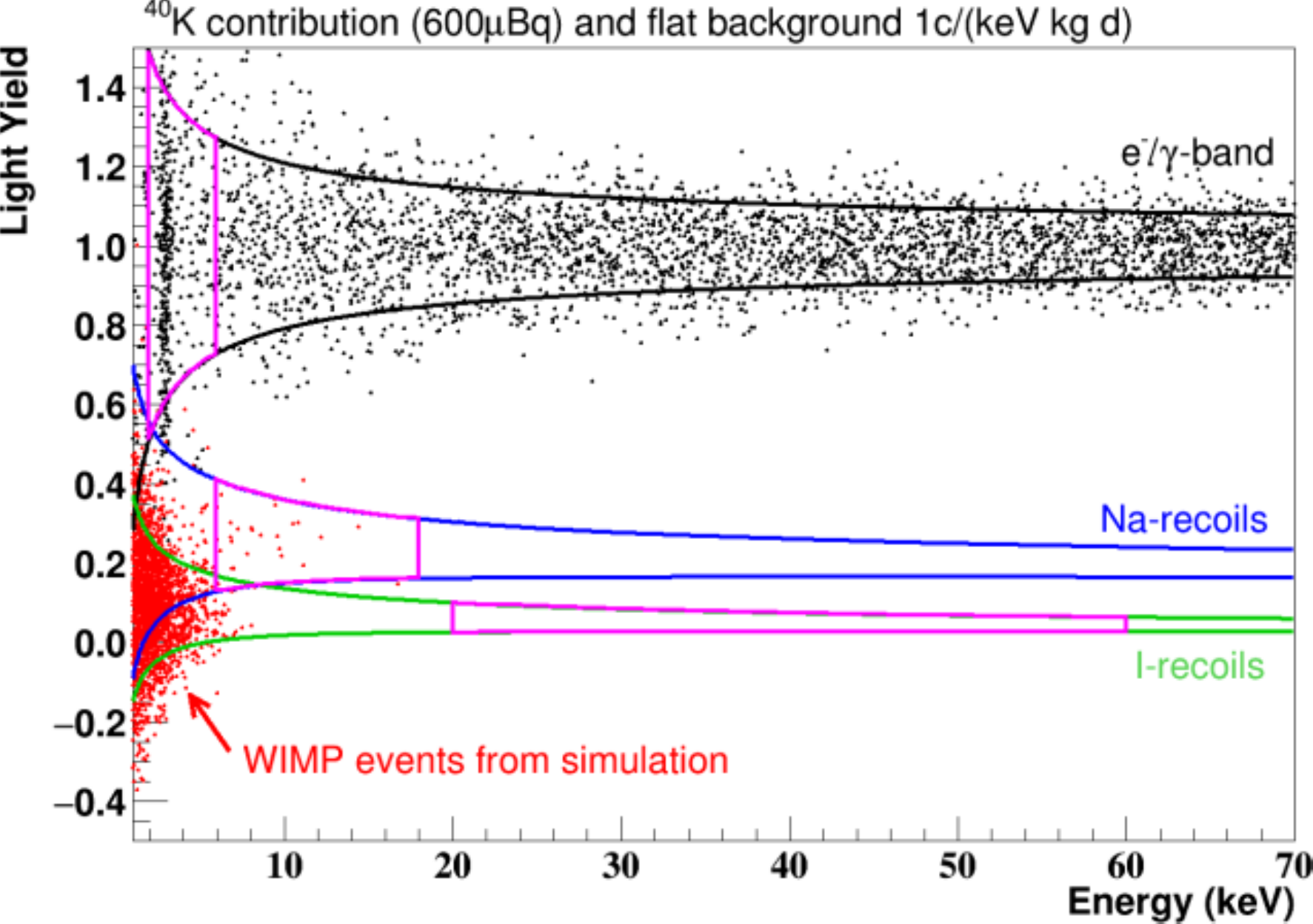}
    \caption{Simulated events for an exposure of 100kg days incl.~e$^-$/$\gamma$-background (black) and DM signal (red). See text for details.}
    \label{fig:LY}
  \end{minipage}
\end{figure}

Fig.~\ref{fig:schema} depicts the detector design. Since the hygroscopic nature of NaI (blue) and its low melting point prevents a direct evaporation of the TES, we instead evaporate it on a carrier crystal (purple).\footnote{For the first prototype (see fig.~\ref{fig:schema}) we used a CdWO$_4$ carrier \textit{connected} with silicon oil to the NaI-crystal.} The light detector is a silicon beaker (black) completely surrounding the target crystal. The beaker shape offers two attractive features: Firstly, it maximizes light collection and secondly it avoids - in combination with the carrier crystal exceeding the diameter of the beaker - any line of sight between the target crystal and non-active surfaces. Such a geometry is mandatory to veto any $\alpha$-induced surface backgrounds \cite{strauss_detector_2015}. 

A subset of the authors recently published results of measurements of CsI \cite{angloher_csi_2016}, performed with a setup comparable to fig.~\ref{fig:schema}. As both, CsI and NaI, belong to the family of alkali halides we are convinced that the experience gained with CsI allows a realistic estimate on the achievable performance of a COSINUS detector. Thereby, a distinct advantage of cryogenic detectors is a very low threshold for nuclear recoils: the design goal of COSINUS is a threshold of 1 keV.\footnote{Currently, the CRESST-II experiment, which is based on the same technology, has word-leading sensitivity for light DM achieved by a nuclear recoil threshold of 0.3 keV \cite{angloher_results_2016}. Furthermore, first prototype measurements indicate that the new CRESST-III detectors reach nuclear recoil thresholds well below 100 eV.} 

Fig. \ref{fig:LY} shows a simulation for an exposure of 100 kg-days in the light yield - energy plane. The light yield is defined by the ratio of light to phonon signal. e$^-$/$\gamma$-events produce most light and, thus, get assigned a light yield of one. Nuclear recoils, instead, have lower light yields indicated by blue and green solid lines for Na and I, respectively.\footnote{The bands for nuclear recoils off Na and I were calculated with the energy-dependent quenching factors of \cite{tretyak_semi-empirical_2010}.} The black events correspond to the background budget reached by the DAMA experiment (flat background of 1 count/keV/kg/day + $^{40}$K activity of 600 $\mu$Bq). To illustrate the discrimination power of a COSINUS detector module we added a DM signal corresponding to the benchmark point shown in fig.~\ref{fig:limit} (m=10 GeV/c$^2$,$\sigma=2\cdot 10^{-4}$pb). We want to stress that this signal reflects the standard scenario of DM particles scattering elastically and coherently off nuclei. The benchmark point was chosen in accordance with the interpretation of the DAMA signal in this scenario as put forward by the authors of \cite{savage_compatibility_2009}. 

As COSINUS detectors measure a phonon signal the threshold is (practically) identical for nuclear and electron recoils. However, for experiments like DAMA only measuring scintillation light, the reduced light output of nuclear recoils (commonly referred to as quenching) has to be considered. This is indicated by the magenta boxes in fig.~\ref{fig:LY} corresponding to electron-equivalent energy of (2-6)keVee which is the energy window where DAMA observes the modulation.
In total, the simulation (for the benchmark point) predicts ~2400 events above threshold with 45\% ranging between energies of (1-2)keV. This simulation clearly points out the two main benefits of a cryogenic calorimeter: low threshold for nuclear recoils and particle discrimination. In conclusion we are convinced that COSINUS detectors will be able to clarify whether the DAMA signals is nuclear recoils or not, even with a moderate exposure of $\mathcal{O}$(10-100 kg-days).    

This conclusion also becomes evident in fig.~\ref{fig:limit} showing the sensitivity (blue band) of COSINUS in comparison to the interpretation of the DAMA signal in the standard scenario of elastic DM nucleus scattering (blue islands \cite{savage_compatibility_2009}). Also shown are the benchmark point and current results from other experiments.

In August 2016 we completed the measurement of the first COSINUS prototype. While its analysis is still ongoing right now, a first event is depicted in fig. \ref{fig:Pulse}. It shows the simultaneous phonon (red) and light (blue) signals\footnote{For the prototype measurement a standard CRESST-like light detector (silicon-on-sapphire disk) was used, instead of the beaker-shaped light detector foreseen for the final COSINUS detector design.} of an e$^-$/$\gamma$-event depositing 240 keV in the detector. To our knowledge this is the first measurement of a NaI-crystal at mK temperatures.

In conclusion, we demonstrate that NaI can be operated as cryogenic calorimeter at mK temperatures. This sheds a positive light on the reachability of the COSINUS performance goals and, therefore, on the possibility to give new insight on the nature of the DAMA signal. 

\section*{Acknowledgements}
\footnotesize{This work was carried out in the frame of the COSINUS R\&D project funded by the INFN (CSN5). In particular, we want to thank the LNGS mechanical workshop team E. Tatananni, A. Rotilio, A. Corsi, and B. Romualdi for continuous and constructive help in the overall set-up construction and M. Guetti for his cryogenic expertise and his constant support.}\newline

\bibliographystyle{iopart-num}
\bibliography{bib.bib}

\providecommand{\newblock}{}
\begin{thebibliography}{1}
\expandafter\ifx\csname url\endcsname\relax
  \def\url#1{{\tt #1}}\fi
\expandafter\ifx\csname urlprefix\endcsname\relax\def\urlprefix{URL }\fi
\providecommand{\eprint}[2][]{\url{#2}}

\bibitem{bernabei_final_2013}
Bernabei R {\em et~al.\/} 2013 {\em EPJ C\/} {\bf 73} 1--11

\bibitem{savage_compatibility_2009}
Savage C {\em et~al.\/} 2009 {\em J. Cos. Astr. Phys.\/} {\bf 2009} 010

\bibitem{angloher_cosinus_2016}
Angloher G {\em et~al.\/} 2016 {\em EPJ C\/} {\bf 76} 441

\bibitem{strauss_detector_2015}
Strauss R {\em et~al.\/} 2015 {\em EPJ C\/} {\bf 75} 1--8

\bibitem{angloher_csi_2016}
Angloher G {\em et~al.\/} 2016 {\em Astropart. Phys\/} {\bf 84} 70--77

\bibitem{angloher_results_2016}
Angloher G {\em et~al.\/} 2016 {\em EPJ C\/} {\bf 76} 1--8

\bibitem{tretyak_semi-empirical_2010}
Tretyak V~I 2010 {\em Astropart. Phys.\/} {\bf 33} 40--53

\end{thebibliography}

\end{document}